\documentclass[twocolumn,amsmath,amssymb,aps,prb]{revtex4}

\usepackage{graphicx}
\usepackage{dcolumn}
\usepackage{bm}
\usepackage{epsf}

\begin{document}
\title{The generalized equation of motion for the orbital dynamics in the presence of current}
\author{Chyh-Hong Chern$^{1}$}
\email{chern@appi.t.u-tokyo.ac.jp}
\author{Naoto Nagaosa$^{1,2,3}$}
\email{nagaosa@appi.t.u-tokyo.ac.jp}

\affiliation{$^1$ERATO-SSS, Department of Applied Physics,
University of Tokyo, Bunkyo-ku, Tokyo 113-8656, Japan. \\
$^2$Correlated Electron Research Center (CERC), National Institute
of Advanced Industrial Science and Technology (AIST),
Tsukuba Central 3, Tsukuba 305-8562, Japan\\
$^4$CREST, Japan Science and Technology Corporation (JST), Saitama,
332-0012, Japan}

\begin{abstract}
The orbital dynamics induced by the charge current is studied
theoretically. The equation of motion for the isospin vector $T^A$
in the SU(N) case is derived in the presence of the current, and is
applied to the cases of $e_g$ (N=2) and $t_{2g}$ (N=3) orbitals. In
spite of the anisotropic transfer integrals between orbitals, the
dynamics is found to be isotropic for $e_g$-orbitals similarly to
the spin case , while it is anisotropic for $t_{2g}$-orbitals. The
implication of this result to the current driven orbital domain wall
motion is discussed.
\end{abstract}
\maketitle

\section{Introduction}
The dynamics of the spins in the presence of the current is an
issue of recent intensive interests
\cite{GMR,SpinValve,Sloncewski,Berger,Bazaliy,Versluijs,Allwood,Yamanouchi,Yamaguchi,SZhang,TataraKohno,SZhang2,Barnes2005PRL}.
Especially it has been predicted theoretically
\cite{Sloncewski,Berger,Bazaliy,SZhang,TataraKohno,SZhang2,Barnes2005PRL}
and experimentally confirmed that the magnetic domain wall (DW)
motion is driven by the spin polarized current in the metallic
ferromagnets \cite{Yamaguchi} and also magnetic semiconductors
\cite{Yamanouchi}. The basic mechanism of this current driven
dynamics is the spin torque due to the current, which is shown to
be related to the Berry phase \cite{Bazaliy}. In addition to the
spin, there is another internal degree of freedom, i.e., orbital
in the strongly correlated electronic systems. Therefore it is
natural and interesting to ask what is the current driven dynamics
of the orbital, which we shall address in this paper. In the
transition metal oxides, there often occurs the orbital ordering
concomitant with the spin ordering. Especially it is related to
the colossal magnetoresistance (CMR) \cite{CMR} in manganese
oxides. One can control this ordering by magnetic/electric field,
and/or the light radiation. Even the current driven spin/orbital
order melting has been observed. On the other hand, an orbital
liquid phase has been proposed for the ferromagnetic metallic
state of manganites \cite{Ishihara1997PRB}. The effect of the
current on this orbital liquid is also an interesting issue.

  There are a few essential differences between
spin and orbital. For the doubly degenerate $e_g$ orbitals, the
SU(2) pseudospin can be defined analogous to the spin. However,
the rotational symmetry is usually broken in this pseudospin space
since the electronic transfer integral depends on the pair of the
orbitals before and after the hopping, and also the spacial
anisotropy affects as the pseudo-magnetic field. For $t_{2g}$
system, there are three degenerate orbitals, and hence we should
define the SU(3) Gell-Mann matrices to represent its orbital
state. There is also an anisotropy in this 8 dimensional order
parameter due to the same reason as mentioned above.

  In this paper, we derive the generic equation of motion of the
SU(N) internal degrees of freedom in the presence of the orbital
current.  N=2 corresponds to the $e_g$ orbitals and N=3 to the
$t_{2g}$ orbitals. Especially, the anisotropy of the DW dynamics
is addressed. Surprisingly, there is no anisotropy for the $e_g$
case while there is for the $t_{2g}$ case. Based on this equation
of motion, we study the DW dynamics between different orbital
orderings. The effect of the current on the orbital liquid is also
mentioned.

\section{CP$^{\text{N}-1}$ Formalism}
In this section, we are only interested in the orbital dynamics.
To be general, we consider a system with $N$ electronic orbital
degeneracy.  The model that we investigate is very similar to the
lattice CP$^{\text{N}-1}$ sigma model with anisotropic coupling
between the nearest-neighbor spinors. In contrary to the
prediction in the band theory, the system is an insulator when it
is $1/N$ filled, namely one electron per unit cell. Because of the
strong on-site repulsive interaction, the double occupancy is not
allowed. Therefore, it costs high energy for electrons to move,
and the spin degrees of freedom is quenched to form some spin
ordering. Due to the complicate spin, orbital, and charge
interplay scheme, it is convenient to use the slave-fermion method
in which we express the electron as $d_{\sigma\gamma i}=h^\dag_i
z^{\text{(t)}}_{\gamma i}z^{\text{(s)}}_{\sigma i}$ where
$h^\dag_i$, $z^{\text{(s)}}_{\gamma i}$, $z^{\text{(t)}}_{\sigma
i}$ are baptized as holon, spinon, and pseudo spinon (for the
orbital) respectively, and the index $i$ denotes the
position\cite{Ishihara1997PRB}.  If holes are introduced into the
system, their mobility leans to frustrate the spin ordering and
thus leads to a new phase which might possess finite conductivity.
Therefore, we consider the following effective Lagrangian
\begin{eqnarray}
L=i\hbar \sum_i (1-\bar{h}_ih_i)\bar{z}_{\alpha i}\dot{z}_{\alpha
i}+\sum_{<ij>}(t^{\alpha\beta}_{ij}\bar{h}_ih_j\bar{z}_{\alpha
i}z_{\beta j}+c.c.) \label{lagrangian}
\end{eqnarray}
where $z$ is short for the spinor $z^{\text{(t)}}$.
The $t^{\alpha\beta}_{ij}$ in Eq.(\ref{lagrangian}) is the transfer
integral which is in general anisotropic because of the symmetry of
the orbitals.  Therefore, the system does not have SU(N) symmetry in
general. To introduce the current, we consider the following mean
field:
\begin{eqnarray}
<\bar{h}_ih_j>=xe^{i\theta_{ij}} \label{current}
\end{eqnarray}
where $x$ denotes the doping concentration, and $\theta_{ij}$ is
the bond current with the relation $\theta_{ij}=-\theta_{ji}$.
 Then, the Lagrangian can be written as
\begin{eqnarray}
L=i\hbar \sum_i (1-x)\bar{z}_{\alpha i}\dot{z}_{\alpha
i}+\sum_{<ij>}(t^{\alpha\beta}_{ij}xe^{i\theta_{ij}}\bar{z}_{\alpha
i}z_{\beta j}+c.c). \label{mean_Lgrngn}
\end{eqnarray}
Note that the constraint $\sum_\alpha | z_{\alpha i}|^2 = 1$ is
imposed, and the lowest energy state within this constraint is
realized in the ground state. The most common state is the orbital
ordered state, which is described as the Bose condensation of $z$,
i.e., $<z_{\alpha i}> \ne 0$. In the present language, it
corresponds to the gauge symmetry breaking. On the other hand, when
the quantum and/or thermal fluctuation is enhanced by frustration
etc., the orbital could remain disordered, i.e., the orbital liquid
state \cite{Ishihara1997PRB}. Then the Lagrangian
Eq.(\ref{mean_Lgrngn}) describes the liquid state without the gauge
symmetry breaking.  In this case, the gauge transformation
\begin{eqnarray}
z_{\alpha i} &\to& e^{-{\rm i} \varphi_i} z_{\alpha i}
\nonumber \\
\bar{z}_{\alpha i} &\to& e^{{\rm i} \varphi_i} \bar{z}_{\alpha i}
\label{gauge}
\end{eqnarray}
is allowed. Given $\theta_{ij}=\vec{r}_{ij}\cdot\vec{j}$, where
$\vec{r}_{ij}=\vec{r}_i-\vec{r}_j$, the local gauge transformation
in Eq.(\ref{gauge}) with $\varphi_i = {\vec r}_i \cdot {\vec j}$
corresponding to the simple shifts of the momentum from $\vec{k}$
to $\vec{k}+\vec{j}$. Therefore, the presence of current does not
affect the state significantly since the effect is canceled by the
gauge transformation.  On the other hand, if $z_{\alpha i}$
represent an orbital ordering, the current couples to the first
order derivative of $z$ in the continuum limit. Define
$a_\mu=i\bar{z}\partial_\mu z$, the second term in
Eq.(\ref{mean_Lgrngn}) can be written as $-j^\mu a_\mu$ in the
continuum limit.  Namely, the current couples to the Berry's phase
connection induced by the electron hopping. Therefore, we expect
some non-trivial effect similar to the spin case.

To derive the equation of motion for the orbital moments, we use the
SU(N) formalism. Introducing the SU(N) structure factors
\begin{eqnarray}
[\lambda^A, \lambda^B]&=&if_{ABC}\lambda^C \\
\{\lambda^A, \lambda^B\}&=&d_{ABC}\lambda^C+g_A\delta^{AB}
\end{eqnarray}
where $\lambda^A_{\alpha\beta}$ are the general SU(N) Gell-Mann
matrices, $[$ $]$ are the commutators, and $\{$ $\}$ are the
anti-commutators. The $f_{ABC}$ is a totally-anti-symmetric
tensor, and $d_{ABC}$ is a totally-symmetric tensor.  Let us
express $t_{ij}^{\alpha\beta}$ in this basis
\begin{eqnarray}
t_{ij}=t^0_{ij}\textbf{1}+t^A_{ij}\lambda^A
\end{eqnarray}
We will only consider the nearest-neighbor hopping:
$t^A_{ij}=t^A_{<ij>}$.  In the rest of the paper,
$t^A_{i,i\pm\hat{k}}$ will be written as $t^A_{k}$, so does
$\theta_{ij}$.  Define the CP$^{\rm{N-1}}$ superspin vector as
\begin{eqnarray}
T^A(i)=\bar{z}_{\alpha i}\lambda^A_{\alpha\beta} z_{\beta i}
\label{isspn_vctr}
\end{eqnarray}
The equation of motion of $T^A(i)$ given by Eq.(\ref{mean_Lgrngn})
can be obtained as \footnote{We also apply the chain rule on the
difference operator.  It differs from the differential operator by
the second order difference. Since the width of the domain wall is
much larger than the atomic scale, our analysis applies.}.
\begin{eqnarray}
\nonumber
\dot{T}^A(i)&=&\frac{xa}{(1-x)\hbar}[-2\cos\theta_{k}f_{ABC}t_{k}^BT^C(i)
\\&-&2\sin\theta_{k}t^0_{x}\Delta_k T^A(i)-\sin\theta_{k}t^B_{k}d_{ABC}\Delta_k T^C(i)
\nonumber\\&&+f_{ABC}t^B_{k}\sin\theta_{k}\jmath^{k}_C(i)]
\label{GLL}
\end{eqnarray}
where the dummy $k$ is summed over $x$, $y$, and $z$ direction, $a$
is the lattice constant, and the orbital current $\vec{\jmath}_C(i)$
is given by
\begin{eqnarray}
\vec{\jmath}_{C}(i)=i(\vec{\Delta}\bar{z}_{\alpha
i}\lambda^C_{\alpha\beta}z_{\beta i}-\bar{z}_{\alpha
i}\lambda^C_{\alpha\beta}\vec{\Delta}z_{\beta i}) \label{bnd_crrnt},
\end{eqnarray}
which is the second order in $\theta$.
Up to the first order in $\theta$, the first term in the
right hand side of the Eq.(\ref{GLL}) is zero provided that
\begin{eqnarray}
t^A_{x}+t^A_{y}+t^A_{z}=0
\end{eqnarray}
which is true for most of the systems that we are interested in.
Consequently, the dominant terms in Eq.(\ref{GLL}) will be the
second and the third ones, which can be simplified as
\begin{eqnarray}
\nonumber
\dot{T}^A(i)&=&-\frac{xa}{(1-x)\hbar}[2\sin\theta_{k}t^0_{x}\Delta_k
T^A(i)\nonumber
\\&+&\sin\theta_{k}t^B_{k}d_{ABC}\Delta_k T^C(i)],\nonumber\\\label{GLL2}
\end{eqnarray}
which is one of the main results in this paper.  Using
Eq.(\ref{GLL2}), we will discuss the orbital DW motion in the
$e_g$ and the $t_{2g}$ systems.

Here some remarks are in order on the mean field approximation
Eq.(\ref{current}) for the Lagrangian Eq.(\ref{lagrangian}).
First, it is noted that the generalized Landau-Lifshitz equation
obtained by Bazaliy {\it et. al}\cite{Bazaliy} can be reproduced
in the present mean field treatment when applied to the spin
problem. As is known, however, there are two mechanisms of
current-induced domain wall motion in ferromagnets
\cite{TataraKohno}. One is the transfer of the spin torque and the
other is the momentum transfer. The latter is due to the backward
scattering of the electrons by the domain wall. Our present mean
field treatment and that in Bazaliy's paper\cite{Bazaliy} take the
former spin torque effect correctly, while the latter momentum
transfer effect is dropped, since the scattering of electrons is
not taken into account. However, the latter effect is usually
small because the width of the domain wall is thicker than the
lattice constant, and we can safely neglect it.

\section{N=2, $e_g$ system}
First, we consider the application on the (La,Sr)MnO (113 or 214)
system \cite{CMR}. Without losing the generality, we show the
result on the 113 system.  The other one can be obtained in a
similar way.

In LaMnO$_3$, the valence of Mn ion is Mn$^{3+}$ with the electronic
configuration $(t_{2g})^3(e_g)^1$.  By doping with Sr, one hole is
introduced to Mn$^{3+}$ and make it to be Mn$^{4+}$.  The transfer
integral between Mn ions depends on the Mn $3d$ and O $2p$ orbitals.
After integrating over the oxygen $p$ orbitals, the effective
hopping between the Mn $d$ orbitals can be obtained.  If we denote the
up state as $d_{3z^2-r^2}$ and the down state as $d_{x^2-y^2}$,
$t_{ij}$ have the following form
\begin{eqnarray}
t_{x}&=&t_0\left(\begin{array}{cc}\frac{1}{4} &
-\frac{\sqrt{3}}{4}
\\ -\frac{\sqrt{3}}{4} &
\frac{3}{4}\end{array}\right)=t_0(\frac{1}{2}\textbf{1}-\frac{\sqrt{3}}{4}\sigma^x-\frac{1}{4}\sigma^z)\nonumber\\
t_{y}&=&t_0\left(\begin{array}{cc}\frac{1}{4} & \frac{\sqrt{3}}{4}
\\ \frac{\sqrt{3}}{4} &
\frac{3}{4}\end{array}\right)=t_0(\frac{1}{2}\textbf{1}+\frac{\sqrt{3}}{4}\sigma^x-\frac{1}{4}\sigma^z)\nonumber\\
t_{z}&=&t_0\left(\begin{array}{cc}1 & 0
\\ 0 &
0\end{array}\right)=t_0(\frac{1}{2}\textbf{1}+\frac{1}{2}\sigma^z)
\end{eqnarray}
where $\sigma^i$ are the Pauli matrices.  For N=2, the pseudospin
moment has the $O(3)$ symmetry given by $T^A=\bar{z}\sigma^Az$.
The consequent equations of motion is given by
\begin{eqnarray}
(\frac{\partial}{\partial
t}+\frac{x}{1-x}\frac{a}{\hbar}t_0\vec{\theta}\cdot\vec{\Delta})T^A(i)=0
\label{GLLeg_1}
\end{eqnarray}
where $\vec{\theta}$ is $(\theta_x,\theta_y,\theta_z)$, which can
be related to the orbital current as
$\vec{v}_o=-\frac{x}{1-x}\frac{a}{\hbar}t_0\vec{\theta}$.  Taking
the continuum limit, Eq.(\ref{GLLeg_1}) becomes
\begin{eqnarray}
(\frac{\partial}{\partial
t}-\vec{v}_o\cdot\vec{\Delta})T^i(\vec{r})=0 \label{GLLeg_2}
\end{eqnarray}
which suggests the solution to be the form
$T^i(\vec{r}+\vec{v}_ot)$.  The result is similar to spin case.
While the spin domain wall moves opposite to the spin
current\cite{Barnes2005PRL}, in our case, the orbital domain wall
also moves opposite to the orbital current.

We can estimate the order of magnitude of the critical current to
drive the orbital DW.  The lattice constant $a$ is about $3{\AA}$.
The transfer integral constant $t_0$ is around $2eV$ in the LSMO
system estimated from the photoemission measurement. If we set $v$
around $1 \text{m/s}$, the critical current can be estimated as
$e/a^3 \sim 6\times 10^{9}$ A/m$^2$, which is roughly the same as
the order of magnitude of that to drive the spin domain wall.

It should be noted that the current only couples to the first
order derivative of $z$. The double exchange term which is given
by the second order derivative is not shown in the equation of
motion. However, the double exchange term plays a role to
stabilize the DW configuration before the current is switched on.

\begin{figure}[htbp]
  \includegraphics[scale=0.4]{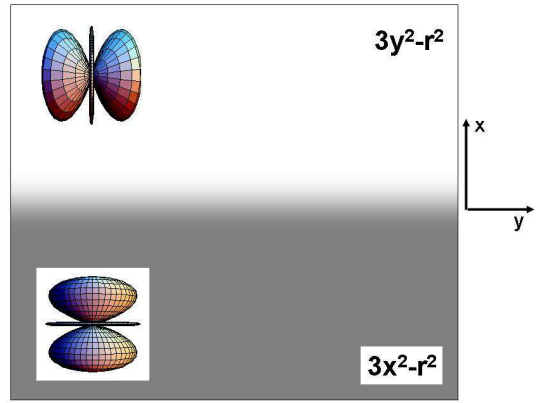}
  \caption{\label{eg_dw}(Color online) An example of the domain structure in the $e_g$ system.  The domain boundary lies on the $y$-axis.}
\end{figure}

There are two orbitals degenerate to $d_{x^2-y^2}$, which are
$d_{y^2-z^2}$ and $d_{z^2-x^2}$.  Similarly, $d_{3y^2-r^2}$ and
$d_{3x^2-r^2}$ are degenerate to $d_{3z^2-r^2}$.  In the Manganite
system, $d_{3z^2-r^2}$ and $d_{x^2-y^2}$ may have different energy
due to slight structural distortion in the unit cell.  Therefore,
in most cases, domain walls are the type of those which separate
two degenerate domains. For example, let's consider the orbital
domain wall separating $3y^2-r^2$ and $3x^2-r^2$.  In
Fig.\ref{eg_dw}, the domain wall sits at $y=0$, and suppose the
current is along the positive $x$ direction.  $3y^2-r^2$ and
$3x^2-r^2$ orbitals are described by the spinors
$(-1/2,-\sqrt{3}/2)$ and $(-1/2,\sqrt{3}/2)$ respectively.  The
configuration given in Fig.\ref{eg_dw} is described by the spinor
field $(\cos\theta(x_i), \sin\theta(x_i))$ where
$\theta(x_i)=2\pi/3(2\cot^{-1}e^{-x_i/w}+1)$, where $w$ is the
width of the domain wall.

Moreover, due to the special properties of SU(2) algebra, the
equation of motion is \emph{isotropic} regardless how
\emph{anisotropic} the transfer integral is.  Therefore, the motion
of orbital domain wall is undistorted as in the spin case.

\section{N=3, $t_{2g}$ system}
Let us consider the $t_{2g}$ systems, for example, in the Vanadate
or Titanate systems.  The $t_{2g}$ orbitals contain three orbits
$d_{xy}$, $d_{yz}$, and $d_{zx}$.  The hopping integral $t_{ij}$
between the Ti$^{3+}$ sites or the V$^{3+}$ ones is given as
\begin{eqnarray}
t_{x}&=&t_0\left(\begin{array}{ccc}1 & 0 & 0
\\ 0 &
0 & 0\\ 0 & 0 & 1\end{array}\right)=t_0(\frac{2}{3}\textbf{1}+\frac{\sqrt{1}}{2}\lambda^3-\frac{1}{2\sqrt{3}}\lambda^8)\nonumber\\
t_{y}&=&t_0\left(\begin{array}{ccc}1 & 0 & 0
\\ 0 &
1 & 0\\ 0 & 0 & 0\end{array}\right)=t_0(\frac{2}{3}\textbf{1}+\frac{1}{\sqrt{3}}\lambda^8) \nonumber\\
t_{z}&=&t_0\left(\begin{array}{ccc}0 & 0 & 0
\\ 0 &
1 & 0\\ 0 & 0 &
1\end{array}\right)=t_0(\frac{2}{3}\textbf{1}-\frac{\sqrt{1}}{2}\lambda^3-\frac{1}{2\sqrt{3}}\lambda^8)
\nonumber\\
\end{eqnarray}
where $\lambda^A$ are SU(3) Gell-Mann matrices with the
normalization condition $Tr(\lambda^A\lambda^B)=2\delta^{AB}$. The
super-spin $T^A$ given by $\bar{z}\lambda^A z$ is normalized
because $\bar{z}z=1$ and $\sum_A
\lambda^A_{\alpha\beta}\lambda^A_{\gamma\delta}=2\delta_{\alpha\delta}\delta_{\beta\gamma}-\frac{2}{3}\delta_{\alpha\beta}\delta_{\gamma\delta}$.

\begin{figure}[htbp]
  \includegraphics[scale=0.4]{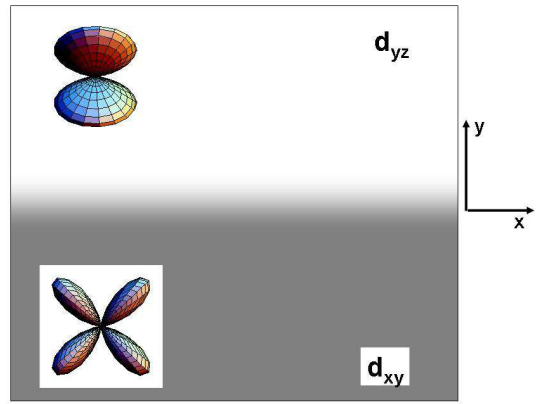}
  \caption{\label{t2g_dw}(Color online) An example of the domain structure in the $t_{2g}$ system.  Only $y-$component of current will move the domain wall.}
\end{figure}

\begin{figure}[htbp]
  \includegraphics[scale=0.4]{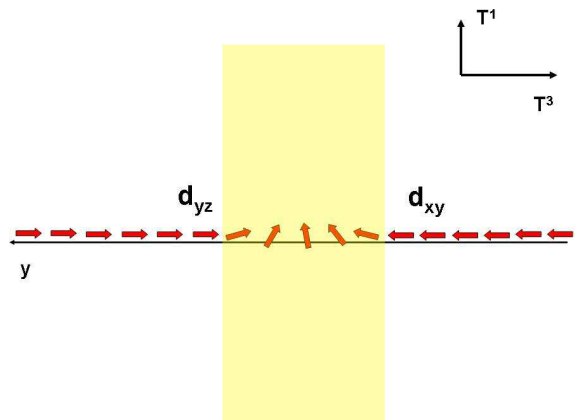}
  \caption{\label{t2g_dw_arrow}(Color online) $d_{xy}-d_{yz}$ domain wall in the superspin space.  The superspin rotates like a XY spin in the superspin space.}
\end{figure}

The equation of motion in this case will be anisotropic because
$d_{ABC}$ is non-trivial in the SU(3) case.  It is inspiring to
work on an example to see how it goes.  Let's consider the
$d_{xy}-d_{yz}$ orbital DW shown in the Fig.\ref{t2g_dw}.  Because
of the orbital symmetry, such DW can only be stabilized along the
$y-$ direction. Similarly, the $d_{xy}-d_{zx}$ DW can only be
stabilized along the $x-$ direction and so on.  Therefore, the
effect of current is anisotropic.  In the absence of current, the
domain wall is stabilized to be
\begin{eqnarray}
z_{\alpha}(\vec{r_i})=\left(\begin{array}{c} \sin\tan^{-1}e^{-y_i/w} \\
\cos\tan^{-1}e^{-y_i/w} \\ 0
\end{array}\right)
\end{eqnarray}
It can be easily seen that it takes no effect if the current is
applied along $x$ or $z$ direction.  The superspin component is
given by
\begin{eqnarray}
T^A(\vec{r})=\left(\begin{array}{c} \text{sech}(y_i/w)\\0\\
\text{tanh}(y_i/w)\\0\\0\\0\\0\\0\\\frac{1}{\sqrt{3}}\end{array}\right)
\end{eqnarray}
When the current is applied along the $y-$ direction,
Eq.(\ref{GLL2}) for each $T^A_k$ are decoupled. For $A=1,2,3$, they
are given by
\begin{eqnarray}
(\frac{\partial}{\partial
t}+\frac{2xt_0}{(1-x)\hbar}\theta_y\frac{\partial}{\partial
y})T^{1,2,3}(\vec{r})=0 \label{t2g_eom_1}
\end{eqnarray}
For $A=4..8$, they are given by
\begin{eqnarray}
(\frac{\partial}{\partial
t}+\frac{xt_0}{(1-x)\hbar}\theta_y\frac{\partial}{\partial
y})T^{4..8}(\vec{r})=0 \label{t2g_eom_2}
\end{eqnarray}
At a glance, we obtain $2$ characteristic drift velocity of the
domain wall.  It is not the case, because $T^4(\vec{r})$ is zero
and $T^8(\vec{r})$ is constant.  Only Eq.(\ref{t2g_eom_1})
determines the motion of the DW .  Furthermore, the
$8-$dimensional super-spin space reduces to be $2-$dimensional as
summarized in Fig.\ref{t2g_dw_arrow}.  $T^1$ moment grows in the
domain wall while $T^3$ moment distinguishes two domains.  In the
presence of current along $y-$direction, the wall velocity is
$|v|=\frac{2xt_0}{(1-x)\hbar}|\theta_y|$.  Other type of domain
structures can be analyzed in a similar way.  As a result, even
though the order parameter in the $t_{2g}$ systems forms an
8-dimensional super-spin space, we can always reduce it to be
2-dimensional because of the anisotropic nature of the system.
Furthermore, the DW moves \emph{without} any distortion just like
the isotropic case.

\section{Discussion and Conclusions}

In this paper, we formulated the orbital dynamics when the spin
degree of freedom is quenched.  We used SU(N) super-spin $T^A$ to
describe the orbital states and obtained the general equation of
motion for it.  We also showed some examples for the SU(2) and
SU(3) cases corresponding $e_g$ and $t_{2g}$ systems,
respectively. In the SU(2) case, the orbital dynamics is very
similar to the spin case: \emph{undistorted and isotropic}.  In
SU(3) case, the DW structure is anisotropic because of the orbital
symmetry.  In addition, the effective super-spin space is
2-dimensional, and the domain wall motion is also
\emph{undistorted}.

Even though the analogy to the spin case can be made, one must be
careful about some crucial differences between the spin and
orbital degrees of freedom. We have estimated the critical current
to drive the domain wall assuming the uniform current flow in the
metallic system, but most of the orbital ordered state is
insulating. This is the most severe restriction when the present
theory is applied to the real systems. An example of the metallic
orbital ordered state is $A$-type antiferromagneitc state with
$x^2-y^2$ orbital ordering in NdSrMnO \cite{CMR}. However it is
insulating along the c-direction, and there is no degeneracy of
the orbitals once the lattice distortion is stabilized. The
ferromagnetic metallic state in LSMO is orbital disordered.
According to the quantum orbital liquid picture
\cite{Ishihara1997PRB}, there is no remarkable current effect on
the orbitals as explained in the Introduction. On the other hand,
when the classical fluctuation of the orbital plays the dominant
role for the orbital disordering, the short range orbital order
can be regarded as the distribution of the domain walls, which
shows the translational motion due to the current as discussed in
this paper. Note however that the current does not induce any
anisotropy in the orbital pseudospin space in the $e_g$ case.

Orbital degrees of freedom is not preserved in the
vacuum or the usual metals, where the correlation effect
is not important, in sharp contrast to the spin.
Therefore the orbital quantum number can not be transmitted
along the long distance and the pseudospin valve phenomenon
is unlikely in the orbital case.

\section{Acknowledgement}
We are grateful for the stimulating discussions with Y. Tokura, Y.
Ogimoto, and S. Murakami. CHC thanks the Fellowship from the ERATO
Tokura Super Spin Structure project. This work is supported by the
NAREGI Grant, Grant-in-Aids from the Ministry of Education,
Culture, Sports, Science and Technology of Japan.

\appendix
\section{$e_g$ orbitals} \label{appendix_A}
There are two class of $e_g$ orbitals, i.e., one is $3x^2-r^2$,
$3y^2-r^2$, $3z^2-r^2$, and the other is $x^2-y^2$, $y^2-z^2$,
and $z^2-x^2$.  They can be described by a two-component spinor. If we use
$(1,0)$ to describe $3z^2-r^2$ and $(0,1)$ for $x^2-y^2$, the other
degenerate orbitals are given by
\begin{eqnarray}
3y^2-r^2 &:& (-1/2,-\sqrt{3}/2) \nonumber \\
3x^2-r^2 &:& (-1/2,\sqrt{3}/2) \nonumber \\
y^2-z^2 &:& (-\sqrt{3}/2, -1/2) \nonumber \\
z^2-x^2 &:& (\sqrt{3}/2, -1/2)
\end{eqnarray}
Their pseudospin moments given by the transformation
$T^A=\bar{z}\frac{\sigma^A}{2}z$ are shown as the following
\begin{eqnarray}
\vec{T}_{3x^2-1}&=&(-\sqrt{3}/2,0,-1/2) \nonumber \\
\vec{T}_{3y^2-1}&=&(\sqrt{3}/2,0,-1/2) \nonumber \\
\vec{T}_{3z^2-1}&=&(0,0,1) \nonumber \\
\vec{T}_{x^2-y^2}&=&(0,0,-1) \nonumber \\
\vec{T}_{y^2-z^2}&=&(\sqrt{3}/2,0,1/2) \nonumber \\
\vec{T}_{x^2-z^2}&=&(-\sqrt{3}/2,0,-1/2)
\end{eqnarray}

\section{Gell-Mann Matrices}
The Gell-Mann matrices used in the text are given explicitly in the
following:
\begin{eqnarray}
\nonumber & & \lambda_1 \!=\! \left(
\begin{array}{cccc}
               0  & 1 & 0   \\
               1  & 0 & 0  \\
               0  & 0 & 0  \end{array} \right), \
 \lambda_2\!=\! \frac{1}{2}\left( \begin{array}{cccc}
               0  & -i & 0 \\
               i  & 0 & 0 \\
               0  & 0 & 0
              \end{array} \right),  \  \\ & &
 \lambda_3 \!=\! \left( \begin{array}{cccc}
                1  & 0 & 0 \\
               0  & -1 & 0 \\
               0  & 0 & 0
               \end{array} \right), \
 \lambda_4 \!=\! \left( \begin{array}{cccc}
                0  & 0 & 1 \\
               0  & 0 & 0 \\
               1  & 0 & 0
               \end{array} \right), \  \nonumber \\
 \nonumber & & \lambda_5 \!=\! \left(
\begin{array}{cccc}
               0  & 0 & -i \\
               0  & 0 & 0 \\
               i  & 0 & 0
               \end{array} \right), \
 \lambda_6\!=\!\left( \begin{array}{cccc}
               0  & 0 & 0 \\
               0  & 0 & 1 \\
               0  & 1 & 0
               \end{array} \right), \ \\ & &
 \lambda_7 \!=\! \left( \begin{array}{cccc}
               0  & 0 & 0 \\
               0  & 0 & -i \\
               0  & i & 0
               \end{array} \right), \
 \lambda_8 \!=\! \frac{1}{\sqrt{3}}\left( \begin{array}{cccc}
               1  & 0 & 0 \\
               0  & 1 & 0 \\
               0  & 0 & -2
               \end{array} \right).
\end{eqnarray}


\end{document}